\documentclass[usenatbib]{mn2e}
\usepackage[utf8]{inputenc}
\usepackage{epsfig,mystyle,amsmath,amssymb,graphicx}
\usepackage{array, xcolor}
\usepackage[colorlinks=true,linkcolor=blue,citecolor=blue]{hyperref}

\usepackage[pass,letterpaper]{geometry}

\pdfoutput=1  % togliere il commento se le figure sono pdf, lasciarlo se sono eps

\usepackage{url}
\usepackage{bm}

\usepackage{tabularx}
\newcolumntype{A}{>{\centering\arraybackslash}X}

\usepackage{graphicx,color}
\usepackage{latexsym,amsmath,amssymb,graphicx,booktabs}
\usepackage{placeins}
\usepackage{mathtools}
\usepackage[normalem]{ulem}

\definecolor{MyBlue}{rgb}{0.15,0.15,0.70}
\definecolor{Dgreen}{rgb}{0,0.7,0.0}

\hypersetup{
colorlinks=true,
citecolor=MyBlue,
linkcolor=MyBlue,
urlcolor=MyBlue
}

\usepackage{amssymb}
\usepackage{amsmath}
\usepackage{amsfonts}
\usepackage{upgreek}
\usepackage{latexsym}
\usepackage{appendix}
\usepackage{dsfont}

\newcommand\ees{\end{eqnarray}}
\newcommand\bees{\begin{eqnarray}}

\usepackage[export]{adjustbox}

\newcommand{\be}{\begin{equation}}
\newcommand{\ee}{\end{equation}}
\newcommand{\bea}{\begin{eqnarray}}
\newcommand{\eea}{\end{eqnarray}}
\newcommand{\dd}{\text{d}}

\newcommand{\baryon}{{\rm b}}

\newcommand{\obh}{$\Omega_{\baryon}h^2$}
\newcommand{\obheq}{\Omega_{\baryon} h^2}

\newcommand{\deu}{${\rm D}$}
\newcommand{\tro}{$^3{\rm He}$}
\newcommand{\qua}{$^4{\rm He}$}

\newcommand{\sep}{$^{7}{\rm Li}$}

\newcommand{\npg}{{$^1$H(n,$\gamma)^2$H}}
\newcommand{\ddn}{{D(d,n)$^3$He}}
\newcommand{\ddp}{{D(d,p)$^3$H}}
\newcommand{\dpg}{{D(p,$\gamma)^3$He}}

\newcommand{\sfac}{$S$--factor}

\begin{document}
%%%%%%%%%%%%%%%%%%% TITLE PAGE %%%%%%%%%%%%%%%%%%%

% Title of the paper, and the short title which is used in the headers.
% Keep the title short and informative.
\title[A tension in cosmology from deuterium?]{A new tension in the cosmological model from primordial deuterium?}

% The list of authors, and the short list which is used in the headers.
% If you need two or more lines of authors, add an extra line using \newauthor
\author[C. Pitrou, A. Coc, J.-P. Uzan, E. Vangioni]{
  Cyril Pitrou,$^{1}$\thanks{pitrou@iap.fr}
  Alain Coc,$^{2}$
  Jean-Philippe Uzan,$^{1}$
  Elisabeth Vangioni$^{1}$
\\
% List of institutions
$^{1}$Institut d'Astrophysique de Paris,
CNRS UMR 7095, 98 bis Bd Arago, 75014 Paris,
France\\
Sorbonne Universit\'e, Institut Lagrange de Paris, 98 bis Bd Arago, 75014 Paris, France\\
$^{2}$IJCLab, CNRS IN2P3, Universit\'e  Paris-Saclay, B\^atiment 104, F-91405 Orsay  Campus France
}

\pagerange{\pageref{firstpage}--\pageref{lastpage}} \pubyear{2020}
\maketitle
\label{firstpage}

\begin{abstract}
Recent measurements of the \dpg\, nuclear reaction cross-section and of the neutron lifetime, along with the reevaluation of the cosmological baryon abundance from cosmic microwave background (CMB) analysis, call for an update of abundance predictions for light elements produced during the big-bang nucleosynthesis (BBN). While considered as a pillar of the hot big-bang model in its early days, BBN constraining power mostly rests on deuterium abundance. We point out a new $\simeq1.8\sigma$-tension on the baryonic density, or equivalently on the D/H abundance, between the value inferred on one hand from the analysis of the primordial abundances of light elements and, on the other hand, from the combination of CMB and baryonic oscillation data. This draws the attention on this sector of the theory and gives us the opportunity to reevaluate the status of BBN in the context of precision cosmology. Finally, this paper presents an upgrade of the BBN code {\tt PRIMAT}.
\end{abstract}


%%%%%%%%%%%%%%%%%%%%%%%%%%%%%%%%%%%%%%%%%%%%%%%%%%
\section*{Introduction} 
%%%%%%%%%%%%%%%%%%%%%%%%%%%%%%%%%%%%%%%%%%%%%%%%%%

Big-Bang nucleosynthesis (BBN) has long been considered as one of the three historical pillars of the cosmological ``Big-Bang'' model, together with the expansion of the universe revealed by the Hubble diagram and the existence of a cosmic microwave background (CMB) of radiation.  In the past decades, the accuracy of the measurements and analysis of these three cosmological probes have drastically improved and were complemented by many other observables, mostly based on the large scale structure of the Universe. As a consequence, the error bars on the cosmological parameters have significantly been improved and, as could have been anticipated, one starts to witness tensions between different probes.

This is in particular the case for the Hubble parameter $H_0$ that is measured to be $69.36\pm0.54$~km/s/Mpc~\footnote{All error bars are stated with $1\sigma$ confidence intervals.} from the global fit of the CMB data~\citep{Planck2018}. This ``low'' value is to be contrasted with the higher value obtained from standard distance ladder, $73.4\pm1.4$~km/s/Mpc~\citep{Reid:2019tiq}, or $73.3\pm1.7$~km/s/Mpc~\citep{Wong:2019kwg} from  strong gravitational lensing effects on quasar systems. In such a situation, one first needs (1) to look for so-far negligible bias in the understanding of each data set, (2) reconsider some hypothesis of the cosmological model, such as the Copernican principle that assumes a spatially homogeneous and isotropic universe, or in this particular case the fluid limit since thin beams~\citep{Clarkson:2011br} do not propagate in the mean Friedmann-Lema\^{\i}tre (FL) spacetime, which can be at the origin of the misinterpretation of the cosmological parameters~\citep{Fleury:2013uqa}. The interpretation of any observation requires to model the propagation of light and is thus tied with the whole cosmological model itself.  To finish (3) one can consider new physics, since here the two discrepant values for $H_0$ correspond to data in the early and late universe; see e.g. \citet{DiValentino:2020zio} for a list of attempts.

As far as BBN is concerned, the theoretical computation rests on the hypothesis of a strictly spatially homogeneous and isotropic FL spacetime, which is thought to be a good approximation in the early radiation dominated universe in which density perturbations are still very small. The microphysics at play is particle and nuclear physics below 100~MeV that can be tested in accelerator. Today, several public (or not) numerical codes are able to predict the abundances of the light elements~\citep{Wagoner1967,Kawano:1992ua,CV17,Parthenope,Parthenope2,AlterBBN,AlterBBN2,Pitrou2018PhysRept,Fie20}.  Prior to WMAP, these predictions depended on two cosmological parameters, the total number of relativistic degrees of freedom (or equivalently the effective number $N_{\rm eff}$ of neutrino families) and the number of baryons per photon $\eta$. This latter quantity is equivalent to specifying the baryon density \obh, a parameter measured by other cosmological probes such as the CMB, with the relation~\citep{Pitrou2018PhysRept}
\bea\label{obhToeta}
\frac{\obheq}{0.0224} &\simeq& \left(\frac{\eta}{6.13197 \times 10^{-10}} \right)\left(\frac{T_{\rm CMB}}{2.7255~{\rm K}}\right)^3\nonumber\\
&&\times \left(\frac{1-1.759\times 10^{-3} \frac{Y_{\rm p}}{0.2471}}{1-1.759\times10^{-3} }\right)\,.
\eea
Prior to WMAP, these parameters were adjustable but they are now determined with high accuracy from the CMB analysis. A first method consists in fixing \obh\, from CMB and $N_{\rm eff}$ from particle physics (thus making BBN a parameter-free model) and assess the agreement between the predicted abundances and the measured ones. Alternatively, we can constrain \obh\ and $N_{\rm eff}$ from BBN (by confronting the predicted abundances which depend on these physical parameters, and the measured ones) and assess the agreement with the values determined by other probes.

Spectroscopic measurements of the abundances of helium-4, deuterium, helium-3 and lithium-7 allow for a comparison with the BBN theoretical predictions. While lithium-7 still exhibits a so-far unexplained discrepancy, see e.g. \citet{Molaro:2009rfa,molaro_vangioni_2009,Fields:2011zzb} for an extended debate, deuterium has been considered as a success of the model due to the agreement of BBN predictions, CMB constraints on \obh\, and observed primitive abundances. Recent measurements~\citep{NatureDPG,Mossa:2020qgj} of one of the key nuclear cross-section drives us to reconsider the robustness of this primordial deuterium success, and more largely of the status of BBN in the standard cosmological model.

%%%%%%%%%%%%%%%%%%%%%%%%%%%
\section{BBN overview}
%%%%%%%%%%%%%%%%%%%%%%%%%%%

BBN predictions consist in abundances of light nuclei (deuterium, helium-3 and -4, lithium-7) that can be compared to spectroscopic measurements and to the trace abundances of heavier nuclei~\citep{Ioc07,Coc:2014oia}, that cannot be measured but may influence the evolution of the first generation of stars. These nuclei are synthetized through nuclear reactions in an expanding universe and can take place only in a narrow window of time during which (1) the thermal bath of the universe has cooled enough for the light atomic nuclei, and foremost deuterium, not to be photo-dissociated, and (2) the density of baryonic matter is high enough for the number of collisions to be large enough. As such it rests on nuclear physics in an expanding homogeneous universe and has two free cosmological parameters, $N_{\rm eff}$ and ${\eta}$.

The predictions reach the percent-level accuracy on helium-4, in complete agreement with its observed value~\citep{Ave20} $Y_{\rm p}=0.2453\pm 0.0034$. Note however that its order of magnitude was initially obtained~\citep{Alpher:1948ve,Alpher:1948gsu} from back-of-the-envelope considerations, because it depends very mildly on $\eta$, and mostly on $\tau_{\rm n}$, $N_{\rm eff}$ (along with the Fermi and Newton constants, $G_{\rm F}$ and $G_{\rm N}$). It was an early and robust prediction of the standard cosmological model~\citep{Peebles:1966rol,Peebles:1966zz} that allowed to claim that only 3 neutrino families existed~\citep{Yang:1978ge}, as was later confirmed by the LEP in 1990. But today, due to its mild dependence on $\eta$ and the accuracy of its measurement, helium-4 is not competitive anymore in our era of precision cosmology to constrain the baryon density. The lithium-7 abundance still exhibits a factor $\sim3$ discrepancy, that is usually discarded with modesty in cosmological studies that never take it into account. The consensus is that it cannot arise from the nuclear sector~\citep{Coc:2014gia,Dav20,IC20}. Helium-3 is less constraining because, (1) since it is both produced and destroyed in stars, the evolution of its abundance in time is not very precise and (2) because there are only few observations in the Galactic disk~\citep{Ban02}. \citet{VangioniFlam:2002sa} have shown that these observations do not allow to set a strong constraint on the primordial baryon density due to the limited understanding of the chemical evolution of this isotope. To finish, deuterium is a very fragile isotope that can only be destroyed after BBN throughout stellar evolution. The most recent recommended observed value provided by \citet{Coo18} is 
\begin{equation}\label{MeasuredDH}
{\rm D}/{\rm H} = (2.527 \pm 0.030) \times 10^{-5}
\end{equation}
 at a redshift $z\sim 2.5-3.1$.

It follows that among all light elements, deuterium is the most constraining since both its observational measurement and its theoretical prediction reach 1\% accuracy. As can been seen from our previous analysis~\citep{Pitrou2018PhysRept}, it requires theoretical predictions and nuclear data to reach the 1\% level so that great care should be paid to nuclear cross-sections affecting deuterium nucleosynthesis.

%While different codes exist~\citep{Wagoner1967,Kawano:1992ua,Parthenope,Parthenope2,AlterBBN,Fie20}, PRIMAT~\citep{Pitrou2018PhysRept} differs slightly in its implementation since it integrates directly differential equations in time, instead of integrating differential equations in the plasma temperature which are obtained by the replacement of $\dd T /\dd t$ derived from the Friedmann equation. 
{\tt PRIMAT}~\citep{Pitrou2018PhysRept} computes directly the weak interaction rates, which interconvert neutrons and protons, including radiative corrections, finite nucleon mass effects, and neutrino spectral distortions, whereas {\tt PArthENoPE}~\citep{Parthenope2} and {\tt AlterBBN} \citep{AlterBBN2} rely on a the fit given in Appendix C of \citet{Serpico:2004gx}. The differential equations governing the evolution of nuclear abundances are integrated in time (as also does {\tt AlterBBN}), which differs from  {\tt PArthENoPE} which integrates equations in terms of the plasma temperature. Since $\dd T/\dd t$ can be obtained from the plasma continuity equation, both methods are of course equivalent. Since its release in 2018, a series of improvements have been included in {\tt PRIMAT} :
\begin{itemize}
\item A refined treatment of neutrino decoupling, including neutrino oscillations and neutrino spectral distortions, has been included by using results from an external neutrino decoupling computation~\citep{Froustey2019,Froustey2020}; 
\item Pair production corrections to nuclear rates that otherwise produce a photon in the final state have been included for the most important reactions~\citep{PitrouPospelov};
\item QED corrections at order $e^3$ have been taken into account in the plasma thermodynamics \citep{Bennett:2019ewm}, whereas previously it was restricted to order $e^2$ corrections. 
\end{itemize}
These three modifications have a very minor impact on $10^5 \times {\rm D}/{\rm H}$ as they shift it by $0.0015$, $-0.0021$ and $-0.0003$ respectively. Only the first modification has a small impact on $Y_{\rm P}$ as it shifts it by $0.00005$, the other two being completely subdominant.

\begin{table*}
\caption{References of the reaction rates in {\tt PRIMAT} 2018~\citep{Pitrou2018PhysRept} and their updated values in {\tt PRIMAT} 2021.} 
\begin{tabular}{lll} 
\hline
\hline
Reaction & {\tt PRIMAT} 2018   &  {\tt PRIMAT} 2021  \\
\hline
\dpg\ &   \citet{Bayes16} & {\bf LUNA} \citet{NatureDPG} \\
$^3$H(d,n)$^4$He &  \citet{Des04} & \citet{Bayes19b}\\
$^3$He(d,p)$^4$He &  \citet{Des04} & \citet{Bayes19a}\\
$^7$Be(n,p)$^7$Li & \citet{Des04} & \citet{Bayes20}\\
$^7$Be(d,p)2$\alpha$ &  \citet{CF88} & \citet{Rij19}\\
\hline
\hline
\end{tabular}
\label{t:netw}
\end{table*}

Also, since the publication of \citet{Pitrou2018PhysRept} there have been a series of updates on the values of the physical parameters. First concerning the cosmology, the value of \obh\, has been revised by the Planck 2018 release~\citep{Planck2018} to
\begin{equation}\label{baryonsCMB}
\obheq =0.02237 \pm 0.00015 \hskip1.5cm \hbox{(CMB)}
\end{equation}
for the CMB alone (instead of the previous $0.02225\pm 0.00016$ from~\citet{Planck2016}), and
\begin{equation}\label{baryonsCMBBAO}
\obheq = 0.02242 \pm 0.00014\hskip1cm \hbox{(CMB+BAO)}
\end{equation}
when combined with baryon acoustic oscillations (BAO) data~\citep{Alam:2016hwk}. The value of the number of effective relativistic degrees of freedom is \citep{Mangano:2005cc,deSalas:2016ztq,Grohs:2017iit,Escudero:2020dfa,Akita:2020szl,Froustey2020,Bennett:2020zkv}
\begin{equation}
N_{\rm eff}=3.044
\end{equation}
for 3 neutrino families\footnote{This recent reference value~\citep{Froustey2020} is lower than the previously admitted $3.046$ of e.g. \citet{Mangano:2005cc} or the improved value $3.045$ of \citet{deSalas:2016ztq}, essentially due to the inclusion of ${\cal O}(e^3)$ QED corrections in the plasma equation of state, following~\citet{Bennett:2019ewm}.}, taking into account the neutrino decoupling physics. This value is very robust and can be understood fully from the adiabatic transfer of averaged oscillations (ATAO) approximation \citep{Froustey2020}. This allows one to show that this prediction is insensitive to the type of neutrino mass hierarchy (normal or inverted) as it depends nearly exclusively on mixing angles. Also, since mixing angles are currently known with rather good precision, the propagation of uncertainty affects $N_{\rm eff}$ with $\pm2\times 10^{-5}$ only.

Then, concerning the microphysics the new neutron decay constant reported by \citet{PDG2020} is
\be\label{taun}
\tau_{\rm n} = 879.4 \pm 0.6\,{\rm s} 
\ee
which is very close to $\tau_{\rm n} = 879.5 \pm 0.8\,{\rm s}$ used in \citet{Pitrou2018PhysRept}, but with an even smaller error bar. It was historically used to bypass the uncertainty about the quark mixing angle $V_{\rm ud}$ and the nucleon axial coupling constant $g_A$ in the prefactor $V^2_{\rm ud}  (1+3 g_A^2)$ which enters the weak interaction rates expressions, thanks to the relation 
\be\label{Directtaun}
\tau_{\rm n} = \frac{2 \pi^3 \hbar}{\lambda_0 G_{\rm F}^2 V_{\rm ud}^2  (1+3 g_{\rm A}^2) (m_{\rm e} c^2)^5}\,,
\ee
with $\lambda_0 \simeq 1.75434$~\citep{2010PhRvC..81c5503C,Pitrou2018PhysRept}. Note that from the recent values\footnote{We use the PDG2020~\citep{PDG2020} value for $g_{\rm A}$, but the PDG2018~\citep{PDG2018} value for $V_{\rm ud}$ since the PDG2020 value for $V_{\rm ud}$ is lower and slightly incompatible with the unitarity of the CKM matrix.} $V_{\rm ud}=0.97420\pm 0.00028$ and $g_{\rm A}=1.2756\pm 0.0013$, we would infer from~\eqref{Directtaun} that $\tau_{\rm n} = 879.4\pm0.5\,{\rm s}$, hence increasing the confidence in the determination~\eqref{taun}.

BBN also notably depends on the value of the Newton constant $G_{\rm N}$ and Fermi constant $G_{\rm F}$ and we rely on their latest CODATA values~\citep{CODATA2018} as well as for all fundamental constants (the sensitivity to these constants has been estimated in works related to the constraints on their possible variations, see e.g. \citet{Coc:2006sx,Uzan:2002vq,Uzan:2010pm}).

Finally, the nuclear network has been updated to take into account the results of new experiments or analyses as summarized in Table~\ref{t:netw}.  None of these changes  brings any relief to the cosmological lithium problem \citep{IC20}.  The reference for the other, unchanged reaction rates, can be found in \citet{Coc12a,Pitrou2018PhysRept}.

The change in the $^7$Be(n,p)$^7$Li rate is mainly responsible for the small decrease of  Li/H. The \citet{Rij19} experiment put the $^7$Be(d,p)2$\alpha$ rate on firmer ground but brings  no change in our  Li/H predictions \citep{CD19}. The rates from the re--analyses of the $^3$H(d,n)$^4$He and   $^3$He(d,p)$^4$He reactions lead to a small change in the \tro/H prediction.  To finish, and that will be the focus of our present analysis, a new reaction rate for the \dpg\ reaction~\citep{NatureDPG} has  recently been published. This is a long awaited and major progress for BBN.

%%%%%%%%%%%%%%%%%%%%%%%%%%%%%%%%
\section{Deuterium nucleosynthesis}
%%%%%%%%%%%%%%%%%%%%%%%%%%%%%%%%

Except for \qua, differences in modern BBN codes are explained by differences in adopted reaction rates. Hence, to compare our results with others, one first needs to compare reaction rates. The production of deuterium mostly depends on 4 nuclear reactions. Deuterium is produced through \npg, the cross-section of which is obtained from an effective field theory computation~\citep{AndoEtAl2006}, reliable at the 1\%-level and in perfect agreement with the existing few experimental data (see e.g. Fig.~1 in \citet{Zakopane}). It is then involved in 3 nuclear reactions  \dpg, \ddn\ and \ddp. These reactions are the main sources of nuclear uncertainty for the prediction of the primordial deuterium abundance. The sensitivity~\citep{CV10} to these reaction rates are
\bea
\frac{\Delta{\rm (D/H)}}{\rm D/H}&=&-0.32\frac{\Delta\langle\sigma{v}\rangle_{\mathrm{D(p,}\gamma)^3\mathrm{He}}}
{\langle\sigma{v}\rangle_{\mathrm{D(p,}\gamma)^3\mathrm{He}}}\label{eq.above}\\
\frac{\Delta{\rm (D/H)}}{\rm D/H}&=&-0.54\frac{\Delta\langle\sigma{v}\rangle_{\mathrm{D(d,n)}^3\mathrm{He}}}{\langle\sigma{v}\rangle_{\mathrm{D(d,n)}^3\mathrm{He}}}
                                                   -0.46\frac{\Delta\langle\sigma{v}\rangle_{\mathrm{D(d,p)}^3\mathrm{H}}}{\langle\sigma{v}\rangle_{\mathrm{D(d,p)}^3\mathrm{H}}}.\nonumber
\label{q:sens}                                                   
\eea
It is clear that a percent accuracy on the predictions, as required by the data, implies to reach a percent level accuracy on the cross-sections. Since none of them have resonances, their determination boils down to the accurate modeling of the slowly varying energy dependent $S$-factor and to a precise determination of their absolute scale.

\subsection{Reaction rate evaluations}

To derive reaction rates and uncertainties,  there are two main approaches in the literature. Either one empirically fits both the energy dependence and scale so as to follow closely the data, or one uses theoretical energy dependences from nuclear physics models and only determine the absolute normalisation. Different approaches have also been considered in the treatment of uncertainties, frequentist versus bayesian with different treatments of systematic uncertainties. However, it has been shown that given the same datasets and fitting functions, those different methods lead to the same 
results for the three reactions. For instance the frequentist \citep{Coc15} and bayesian \citet{Bayes16} \dpg\ rates are
almost identical (see next section). Similarly, Eqs. (3.49)--(3.51) from \citet{Ser04} were tested in \citet{Coc15}
leading to very similar rates and uncertainties. Hence, the differences in reaction rates obtained from different
groups come from the selection of datasets and the choice of fitting function.   

A major difficulty, for those three reactions is that only a few 
experimental datasets were obtained by  precision experiments dedicated to BBN (e.g. \citep{NatureDPG} for \dpg\
or \citet{Leo06} for \ddn\ and \ddp). Many datasets lack sufficient documentation concerning the scale (systematic)
error. This is the main criteria used by \citet{Coc15,Bayes16,Bayes17} to exclude datasets.
In several cases, the scale error is not evaluated or only the combined, statistical and systematic 
uncertainties are given so that the corresponding datasets are also put aside. Details on this selection
are given in \citet{Coc15,Bayes16,Bayes17}. 

The other issue concerns the \sfac\ fitting function. One option is to use polynomial (e.g. \citet{Ser04,Cyb04}) or 
splines  (e.g. \citet{Nol00}), but choosing the correct  polynomial degree is difficult. A higher degree provides
a better fit, but can introduce artificial structures. This is why many evaluations introduce some phenomenological
(e.g. R--matrix in \citet{Des04} or Potential Model from \citet{NACRE2}) or even theoretical prejudices (e.g. {\it ab initio} 
model of \citet{Nef11}). 
In previous works, (e.g. \citet{Pitrou2018PhysRept}), we had chosen this latter option, since {\it ab initio} \sfac{s} 
were available  for the three reactions \citep{Mar05,Ara11}.    

Finally, the \dpg, \ddn\ and \ddp\  adopted rates in \cite{Pitrou2018PhysRept} result from bayesian analyses 
\citep{Bayes16,Bayes17}.  They have
the advantage of not being limited to gaussian distributions and to be able to take into account systematic 
uncertainties in a simple way \citep{Bayes16} (see also \citet{Bayes19b,Bayes19a,Bayes20} concerning other 
reactions). However, note that for the \dpg\ rate we use the latest LUNA rate by \citet{NatureDPG} (see below).

\subsection{The \dpg\ rate}

This reaction rate has long been a subject of controversy. As displayed in Fig. 23 of \citet{Pitrou2018PhysRept} (updated in Fig.~\ref{f:sfac} below),  there was a scarcity of experimental data in the region of interest for BBN.

In their evaluations, \citet{Coc15} and \citet{Bayes16} used  the theoretical \sfac\ from \citet{Mar05} re-normalized (e.g. a factor of 0.9900$\pm$0.0368 in \citet{Coc15}) to a selection of experimental data.  Other authors~\citep{Cyb04,Des04}  have preferred the alternative option that follows closely the experimental data points,  resulting in a lower \sfac\ at BBN energies, mainly driven by the \citet{Ma97} data (see Fig.~\ref{f:sfac}).
The widely used  NACRE--II \citep{NACRE2} compilation relies for this reaction on a potential model, adjusted to experimental data, but gives little details.  
The  NACRE--II \citep{NACRE2} compilation was designed to be conservative i.e. their \sfac\ limits were supposed to encompass almost all existing data, in order to be sure that the real \sfac\ is within the limits. The  problem became more acute with the publication of an improved theoretical \sfac\ by \citet{Mar16},  lying above the previous one  of \citet{Mar05}; see Fig.~\ref{f:sfac}.  Very recently, this cross--section, of the most important reaction for deuterium destruction, has been measured, first at   the Joi{\v{z}}ef Stefan Institute of Ljubljana by \citet{Tis19}, then at the LUNA, Gran Sasso underground laboratory \citep{NatureDPG} (see Fig.~\ref{f:sfac}). Those experiments explored the energy range relevant to BBN.  In particular, the LUNA data points span the range $E_\mathrm{cm}$ = 32--263~keV, and have very small error bars.

\begin{figure*}
%\begin{center}
\includegraphics[width=0.85\textwidth]{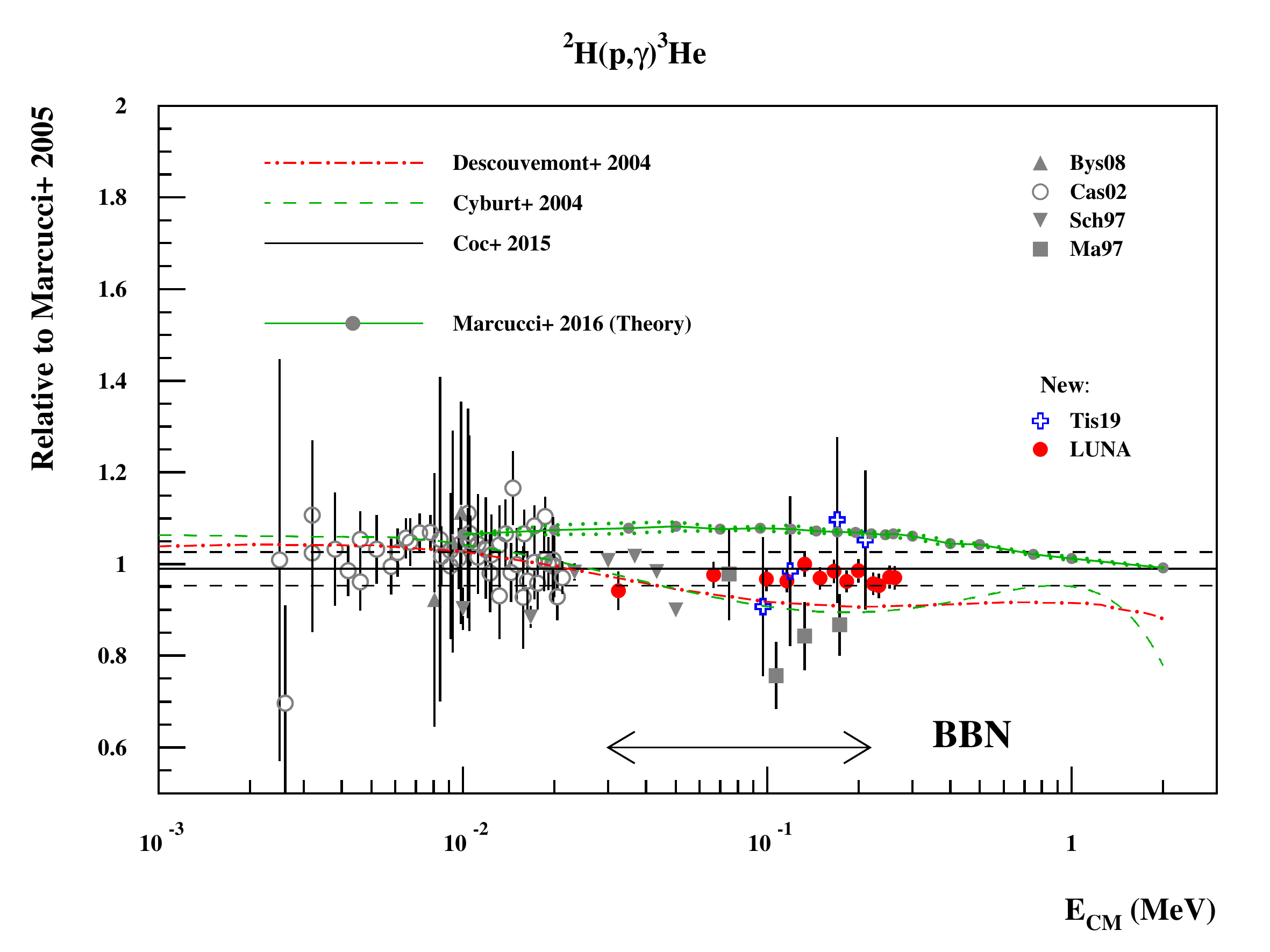}
%\end{center}
\caption{Theoretical and experimental \sfac, normalized to the \citet{Mar05} theoretical one. 
Data in grey point "Bys08" \citep{Bys08},  "Ca02" \citep{Cas02},  "Sch97" \citep{Sch97} and  "Ma97" \citep{Ma97}, are those used in our previous calculations~\citep{Coc15,Bayes16,Pitrou2018PhysRept}. 
Blue points \citep{Tis19} and red points (LUNA \citet{NatureDPG}) are new. Compared to Fig. 23 of \citet{Pitrou2018PhysRept}, only datasets that were used in \citet{Coc15,Bayes16} are displayed. Curves are the \sfac\ used in previous BBN calculations (see text).} 
\label{f:sfac}
\end{figure*}  

\vspace{0.2cm}
From these experiments, one can deduce that
\vspace{-0.2cm}
\begin{itemize}
\item the LUNA data \citep{NatureDPG} confirm, in the BBN range, the energy dependence and magnitude of the \sfac\ calculated by \citet{Mar05} (Fig.~\ref{f:sfac}), 
\item the new data \citep{Tis19,NatureDPG} do not confirm the low \sfac\ from \citet{Ma97} that, previously drove down the fitted \sfac{s} \citep{Des04,Cyb04}, 
\item does not confirm the higher theoretical \sfac\ from \citet{Mar16}, and
\item the LUNA data lies in between the \sfac\ limits derived by \citet{Coc15} and, not shown on the Figure, those subsequently obtained by a more sophisticated (Bayesian) analysis by \citet{Bayes16}.
\end{itemize}

\begin{figure}
%\begin{center}
\includegraphics[width=\columnwidth]{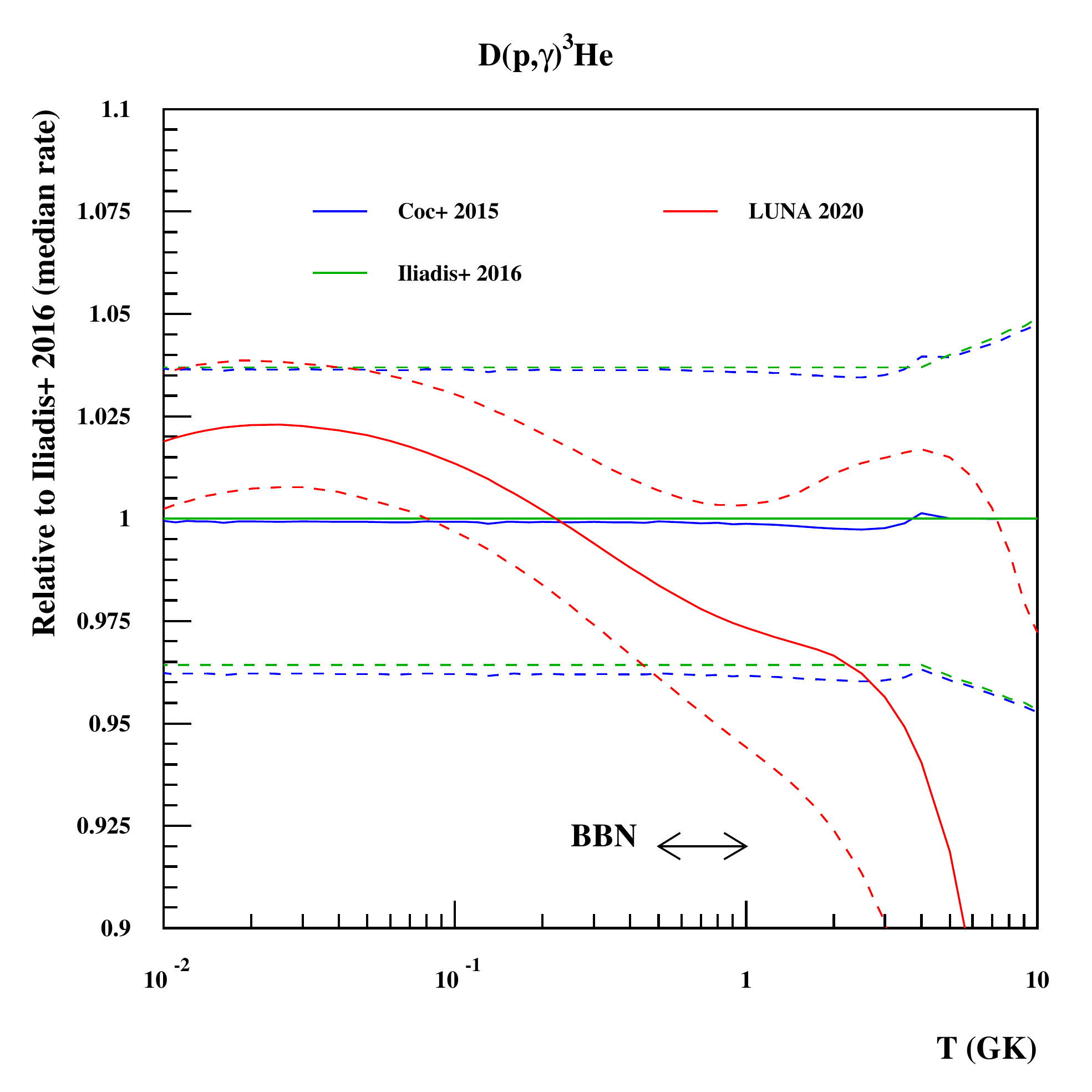}
%\end{center}
\caption{LUNA reaction rates  \citep{NatureDPG} and uncertainties compared to the ones previously used \citep{Coc15,Bayes16}.
Rates labelled Coc+ 2015 are deduced from corresponding \sfac{s} in Fig.~\ref{f:sfac}, those
labelled Iliadis+ 2016 are those used by \citet{Pitrou2018PhysRept}.}
\label{DPGfig}
\end{figure}  

Consequently, the rate \citep{Bayes16} used  by \citet{Pitrou2018PhysRept} will need only minor revision (Moscoso {\em et al.}, in preparation), and confirm the deuterium tension, first observed by \citet{Coc15,Pitrou2018PhysRept}. 
Indeed, Fig.~\ref{DPGfig} compares the rate \citet{Bayes16} previously used by \citet{Pitrou2018PhysRept} with the new rates recently derived from Eq.~(2) and (3) in \citet{NatureDPG}, in agreement with their Table~1.  
In the BBN temperature range, the new rate is mostly within the limits of the previously adopted ones. 
\citet{Fie20} use the \dpg\ rate from NACRE--II \citep{NACRE2} as a baseline, but also consider those from \citet{Coc15} (very close to the one \citet{Bayes16} used in {\tt PRIMAT}; see Fig.~\ref{DPGfig}) and the high  theoretical rate of \citet{Mar16}.  Few details are given in NACRE--II on the evaluation of the \dpg\ rate, but it is found to be significantly lower than the one used in {\tt PRIMAT} and  has wider limits.  Hence, its use by \citet{Fie20} is expected to lead to a higher D/H prediction.  

To take into account the new experimental data we use  the \citet{NatureDPG} rate,  derived from their
Eq.~(2) and (3), making comparison with other works easier. 

\subsection{The \ddn\ and \ddp\ rates}

\begin{figure}
%\begin{center}
\includegraphics[width=\columnwidth]{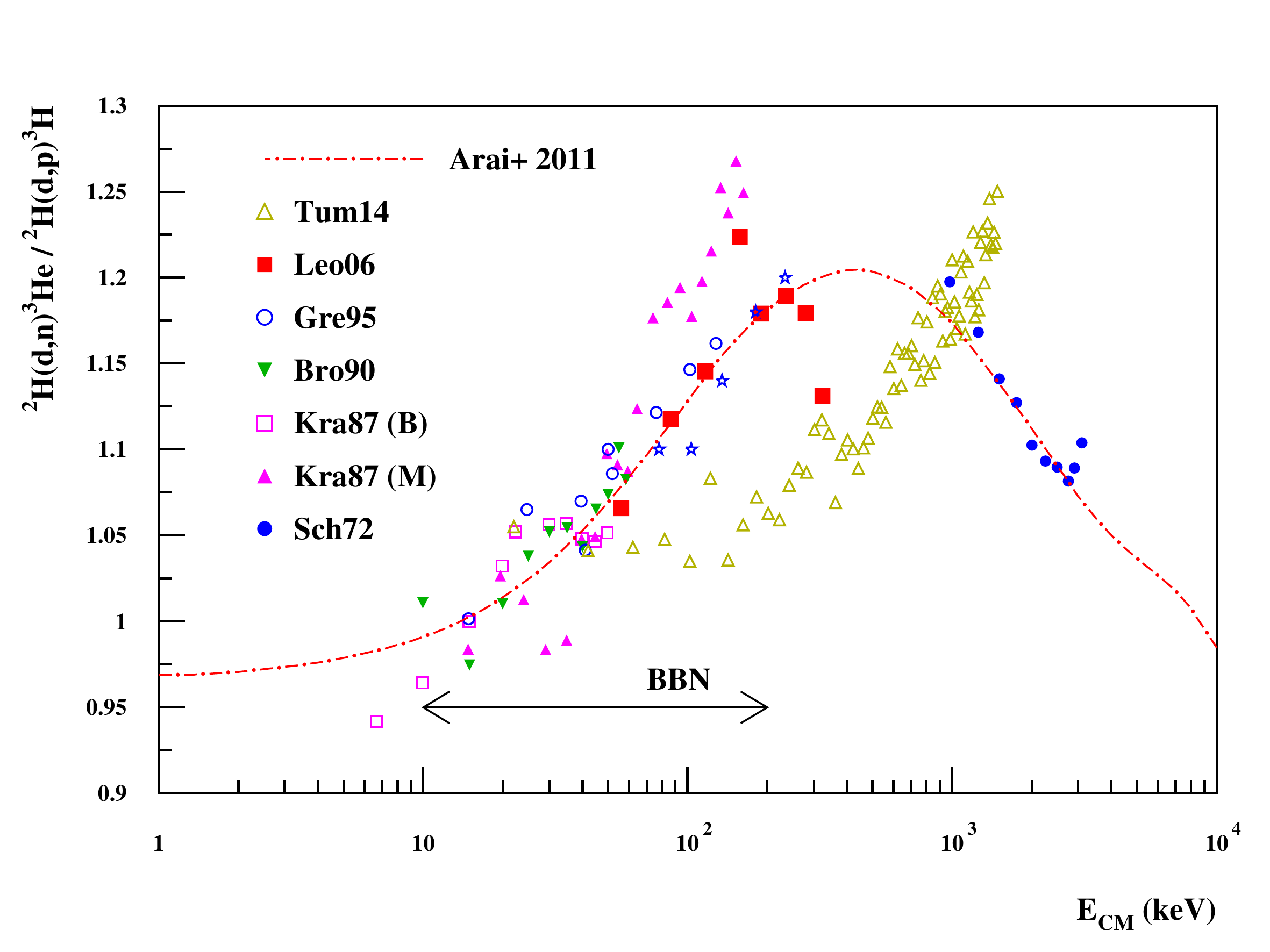}
%\end{center}
\caption{Ratio of the \ddn\ over \ddp\ cross sections.  BBN, recent experimental data from direct measurements 
(``Kra87" \citep{Kra87}, ``Bro90" \citep{Bro90}, ``Gre95" \citep{Gre95}, ``Leo06" \citep{Leo06} and ``Sch72" \citep{Sch72}) 
follows the theoretical predictions of \citet{Ara11}).  The indirect data from \citet{Tum14} follows a different trend. }
\label{f:ddnp}
\end{figure}

As reminded in Eq.~(\ref{eq.above}), two other reactions are important for deuterium destruction:  \ddn\ and \ddp. For these reactions, {\tt PRIMAT} relies on the rates evaluated by \citet{Bayes17}, based  on the theoretical, {\it ab initio} energy dependences from \citet{Ara11} re-normalized to a selection of experimental data, using bayesian techniques. \citet{Fie20}  use instead the NACRE--II rates based on a DWBA model adjusted to experimental data. However, as for \dpg, few details are available in  NACRE--II on the evaluation of experimental data, and rate uncertainties. 
Contrary to the \dpg\ reaction, several recent experimental studies have investigated both \ddn\ and \ddp\   
cross sections at BBN energies \citep{Kra87,Bro90, Gre95,Leo06}. These are all direct measurements that
are in good agreement with the theoretical cross section obtained by \citet{Ara11} as can be seen in Fig.~\ref{f:ddnp}.
It displays the ratio of \ddn\ over \ddp\ \sfac{s}, allowing to evaluate the coherence of the data because this ratio 
is essentially governed by the Coulomb interaction, and as such is weakly dependent of the nuclear model.
The theoretical curve \citep{Ara11}
(not a fit) reproduces the directly measured data, including the \citet{Sch72} at high energy, above the BBN range.
Reaction rates based on a re--normalization of the \citet{Ara11} \sfac\ to the  experimental data of \citet{Kra87,Bro90, Gre95,Leo06} 
were obtained by \citet{Coc15} and \citet{Bayes17}  analyses. 
These four experimental studies  were selected because they all provide both statistical
and systematic uncertainties.  
In particular the most recent direct experiment \citep{Leo06} provides an error matrix and quote 
a scale error as low as 2\%$\pm$1\%. 
Both uncertainties were considered separately by \citet{Coc15}, using a classical 
analysis, and by \citet{Bayes17} with a Bayesian analysis that, in particular treats systematic uncertainties
as priors. Resulting reaction rates were found to differ by less than 0.2\% and we adopt the \citet{Bayes17}
rate.
The  \ddn\ and \ddp\  rates used in the LUNA BBN calculations \citep{NatureDPG} are
updated from the \citet{Parthenope2,Ser04} evaluation including a minor contribution from
the new data  \citep{Tum14} obtained by the (indirect) Trojan Horse Method.
The main differences with the \citet{Bayes17} analysis is that the latter applies stricter selection  criteria on 
experimental data (e.g. only direct measurements with
evaluation of systematic uncertainties) and uses theoretical guidance instead of polynomials.

In conclusion, our  BBN results  \citep{Coc15,Pitrou2018PhysRept,IC20} for D/H are in general lower than others, because we use different reaction rates for \dpg\ (previously \citet{Bayes16}, but here, replaced by LUNA \citep{NatureDPG}), 
\ddn, and \ddp\ (\citet{Bayes17}). In these evaluationsd \citep{Bayes16,Bayes17}, first, only experimental datasets whose error budget (statistical {\em and} systematics) is available, are adopted. Next, whenever possible, theoretical guidance is considered.
% (\citet{Mar05} for \dpg, and \citet{Ara11} for \ddn\ and \ddp).
Other works may use smooth polynomial fits to  the data, which is, in principle, another reasonable option. Finally, our
adopted rates  are obtained using Bayesian techniques because they allow for a rigorous inclusion of statistical {\em and systematic} sources of uncertainties. These choices have the advantage of being fully documented and simply stated.
However, for this work, we use provisionally the \citet{NatureDPG} rate.

%%%%%%%%%%%%%%%%%%%%%%%%%%%%%%%%%%%%%%%%%%%%%%%%%%%%%%%%
\begin{table*}%[htbp!] 
\caption{\label{t:hlix} Predicted abundances compared to observations.}
\begin{center}
\begin{tabularx}{0.88\textwidth}{lccc}
\toprule
& Observations & \citet{Pitrou2018PhysRept} $\tau_{\rm n} = 879.5(8)\, {\rm  s}$, &  {\bf This work} $\tau_{\rm n} = 879.4(6)\, {\rm  s}$, \\
&&$100h^2 \Omega_b = 2.2250\,(\pm 0.016)\, ({\rm e})$&$100 h^2 \Omega_b = 2.242\,(\pm 0.014)\,({\rm f})$\\
\midrule 
$Y_{\rm P}$    & 0.2453$\pm$0.0034(a)  & {0.24709}$\pm${0.00018} & {\bf 0.24721}$\pm${\bf 0.00014}\\
\deu/H   ($ \times10^{-5})$ & 2.527$\pm$0.030 (b)& {2.460}$\pm${0.046}    &  {\bf 2.439}$\pm${\bf 0.037}   \\
${}^3{\rm He}/{\rm H}$    ($ \times10^{-5}$)  & $<$1.1$\pm$0.2 (c)&  {1.074}$\pm${0.026}  &  {\bf 1.039}$\pm${\bf 0.014}\\
\sep/H ($\times10^{-10}$)  &   1.58$^{+0.35}_{-0.28}$ (d) & {5.627}$\pm${0.259} &  {\bf 5.464}$\pm${\bf 0.220}\\
\bottomrule
\end{tabularx}\\%}
(a) \citet{Ave20}, (b) \citet{Coo18}, (c)\citet{Ban02}, (d) \citet{Sbo10}, (e) \citet{Planck2016}, (f) {\rm CMB+BAO}, \citet{Planck2018}
\end{center}
\label{TableAbundances}
\end{table*}

%%%%%%%%%%%%%%%%%%%%%%%%%%%%%%%%%%%%%%%%%%%%%%%%%%%%
\section{Constraints on cosmological parameters from BBN}
%%%%%%%%%%%%%%%%%%%%%%%%%%%%%%%%%%%%%%%%%%%%%%%%%%%%

As mentioned above, there are two equivalent ways to look at the data. Either, we use BBN to constrain the only free cosmological parameter that affects the abundances, i.e. the baryonic density, and we then compare this measurement to the one by Planck~\citep{Planck2018} (CMB or CMB+BAO), or we fix the baryonic density to its value determined by CMB analysis and compare the predictions of BBN under that hypothesis to spectroscopic data.

\begin{figure}
%\begin{center}
  \includegraphics[width=0.99\columnwidth]{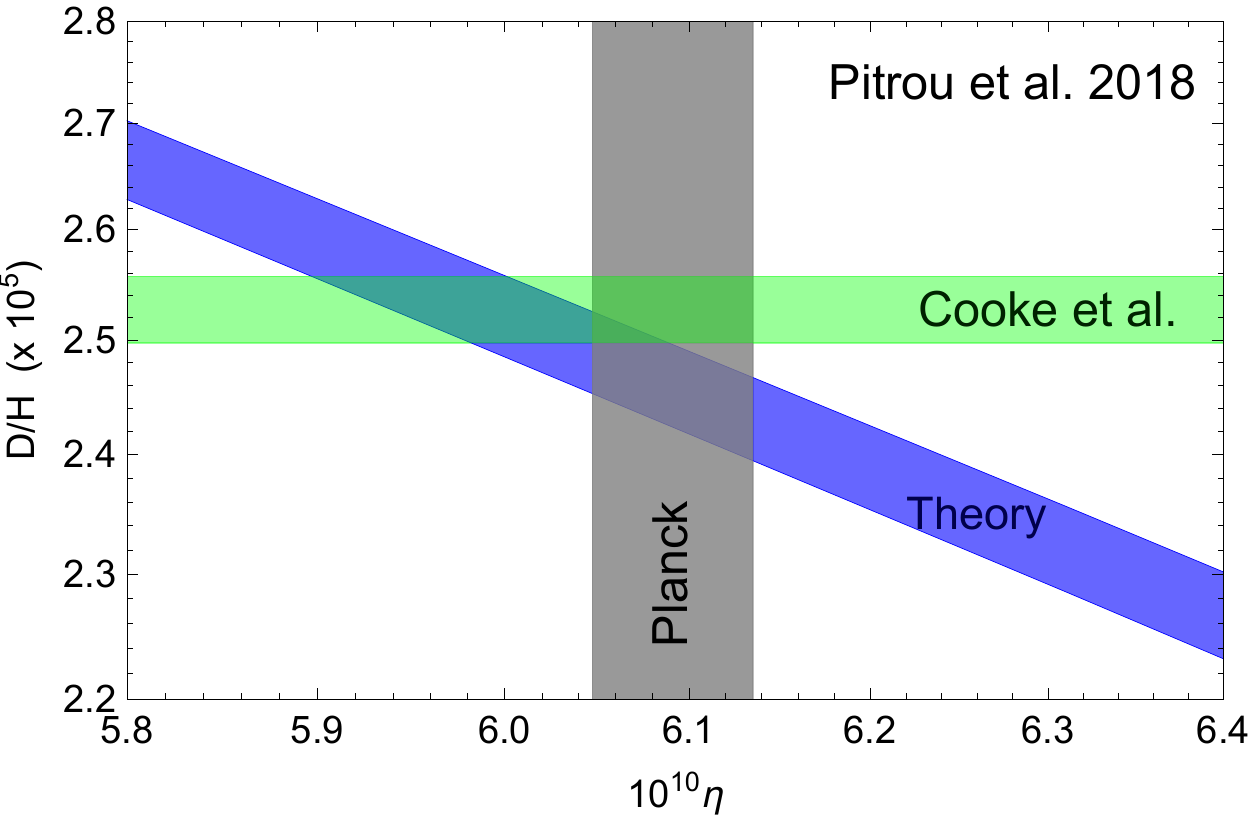}\\
  \includegraphics[width=0.99\columnwidth]{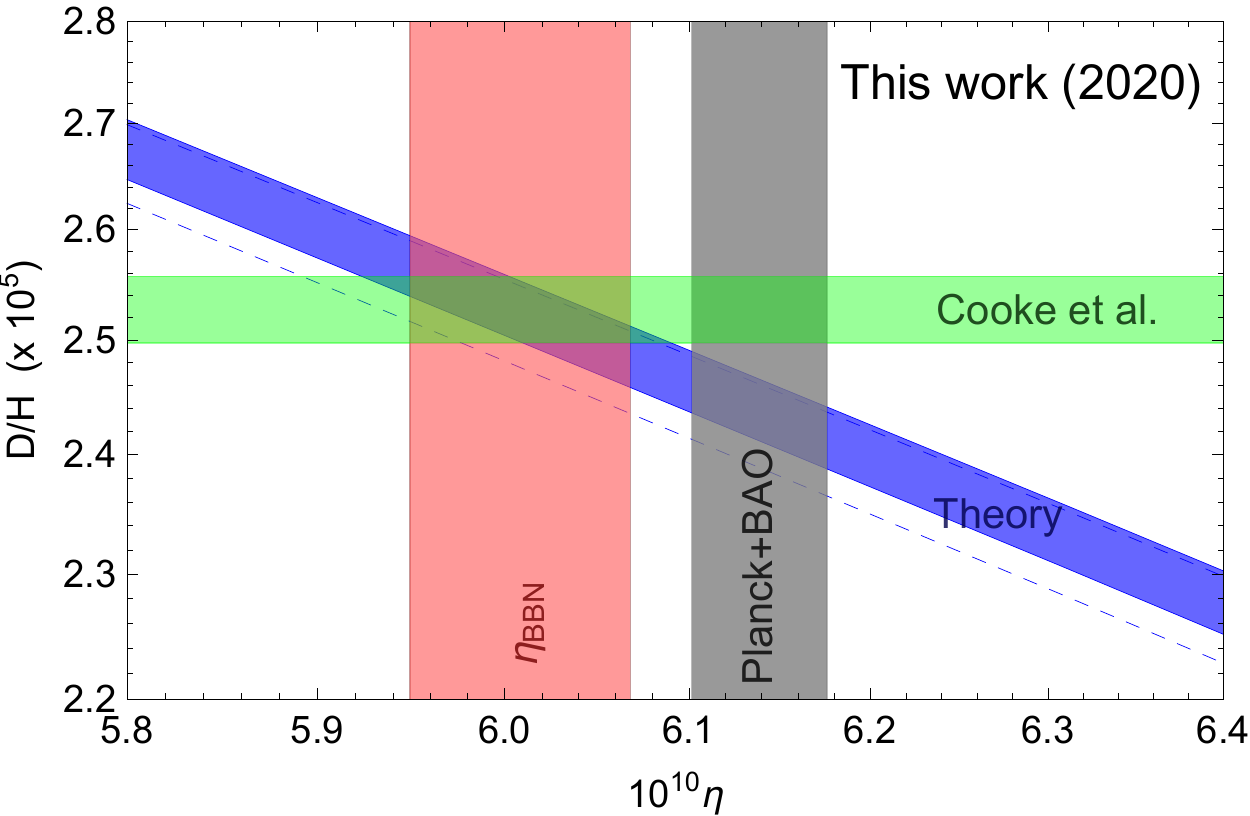}\\
  \includegraphics[width=0.97\columnwidth]{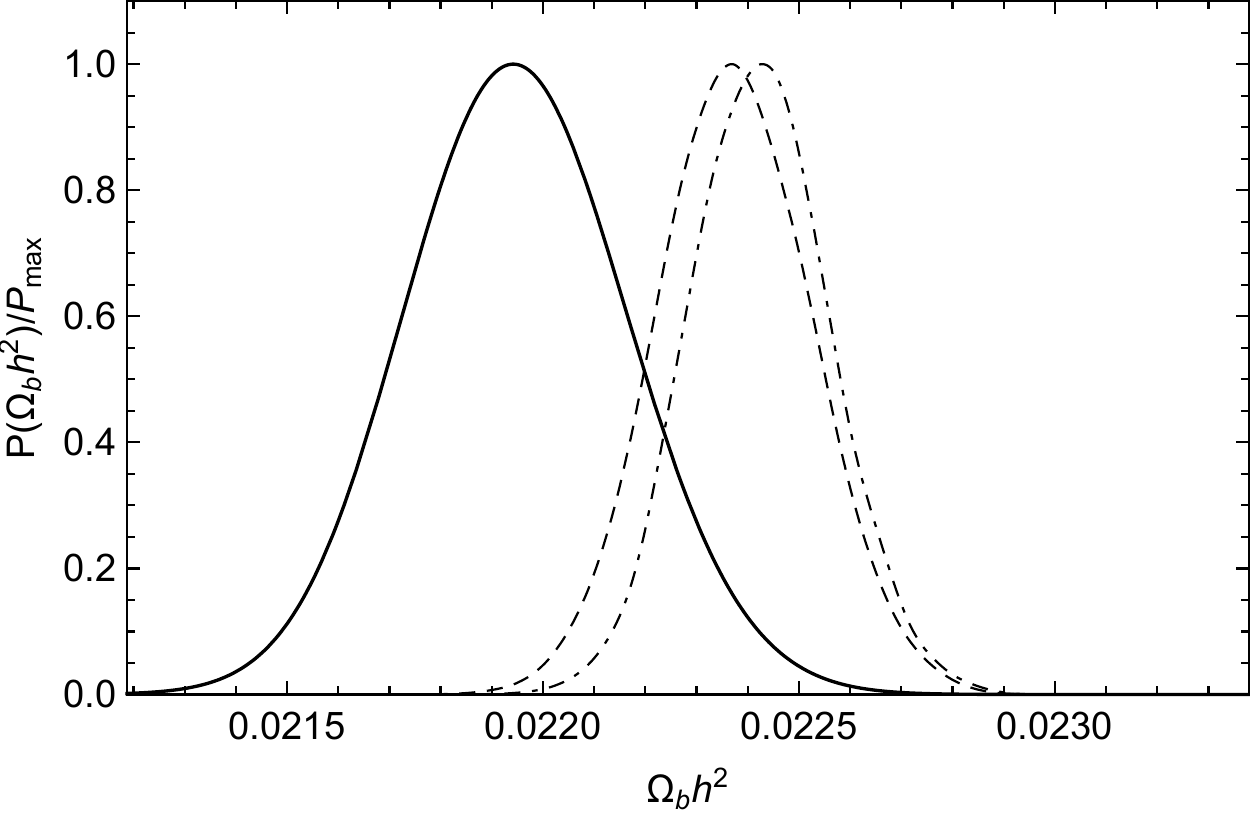}
%\end{center}
\caption{{\it Top:} D/H theoretical prediction (in blue), observation (in green) from \citet{Coo18}, and baryon abundance constraints from CMB (in gray), as reported in \citet{Pitrou2018PhysRept}. All ranges displayed are within $1\sigma$ standard deviation. {\it Middle:} Same quantities but the baryon density is updated from the CMB+BAO constraint by \citet{Planck2018}, and with the D/H theoretical predictions of this work. The dashed blue lines correspond to the theoretical range determined when using the \dpg\, rate of \citet{Bayes16} instead of the recent LUNA rate~\citep{NatureDPG}. {\it Bottom:} Posterior distribution of baryon density from BBN (this work) in solid line, from CMB only in dashed line, and from CMB+BAO in dot-dashed line (both from \citet{Planck2018}). The correspondence between $\eta$ and \obh\ is given by Eq.~\eqref{obhToeta}.}
\label{DHzoom}
\end{figure}

Figure~\ref{DHzoom} summarizes the predictions for BBN deuterium from the present analysis
[using \citep{NatureDPG}  for the \dpg\ rate, and \citet{Bayes17} for the \ddn\ and \ddp\ rates] 
and the previous one by \citet{Pitrou2018PhysRept}, as well as the CMB constraint on $\eta$ and the data by \citet{Coo18}.

In the first approach, we use BBN theory and spectroscopic observations to determine $\eta$, assuming that $N_{\rm eff}$ is fixed from particle physics, and compare to its CMB value by Planck~\citep{Planck2018}. Using the method described in section 6.2 of \citet{Pitrou2018PhysRept}, we estimate the posterior distribution of \obh, given the observational constraints on \qua\
% (as in \citet{Pitrou2018PhysRept} we use the constraints from \citet{Ave15})
and on \deu. The posteriors for CMB or BBN determinations of \obh\ are depicted on Fig.~\ref{DHzoom}, and we obtain for BBN only
\be
\obheq = 0.02195 \pm 0.00022\,.
\ee
This is a $1.6\sigma$ tension with CMB~\eqref{baryonsCMB} and $1.84\sigma$ tension with CMB+BAO~\eqref{baryonsCMBBAO}. The tension is higher when BAO are included, which is in general the case when more data are considered. Note also that BAO favour baryons compared to dark matter in the analysis.

Equivalently, the same analysis can be performed by assuming that the baryon density is determined from CMB+BAO~\citep{Planck2018}, and predict the theoretical expectation for the deuterium abundance. When estimating the theoretical uncertainty with a Monte-Carlo method, we vary on the uncertainty of nuclear rates, on the neutron lifetime, but also on the baryon abundance according to the CMB+BAO posterior. We then find the theoretical expectation
\be
\rm (D/H) = (2.439 \pm 0.037)\times 10^{-5}\,.
\ee
Again, this has a $1.84\sigma$ tension with the measured value~(\ref{MeasuredDH}). This is expected since these two methods are different ways of doing the same thing.

%%%%%%%%%%%%%%%%%%%%
\section*{Discussion and perspectives}
%%%%%%%%%%%%%%%%%%%%

BBN theory has long been considered as a standard pillar of the big-bang model, despite the long-standing lithium-7 problem. In the current era of precision cosmology, its constraining power mostly rests on the prediction of the deuterium abundance since the accuracy of helium-4 data, and its mild dependence on the baryon density, do not make it a competitive probe anymore. As we argued, the agreement between data, theoretical BBN predictions and CMB constraints on the baryonic density requires to control the accuracy of the BBN computation at the percent level.
%Today, the main difference between all existing codes lies almost entirely on the handling and the choice of strategy concerning the nuclear physics.

First, it follows that nuclear data are the crux in this debate. %As explained and quantitatively evaluated,
All existing codes  differ from the difference of choices on the modelisation of the nuclear cross-sections, and not on weak rates since they differ by less than 0.2\% between e.g. {\tt PRIMAT} and {\tt PArthENoPE}~\citep{Parthenope}. It is important to control their accuracy at least at the percent level and to take into account the latest data. The recent release of the LUNA data confirms the \sfac\ and rate previously used in {\tt PRIMAT} 2018~\citep{Pitrou2018PhysRept}. We updated it to fully take into account these data and the code now also includes a series of refinements described in this article. Finally, because of the importance of the d+d  reaction rates, in particular, the \ddn\ one, further investigations are needed to reconcile Trojan Horse results \citep{Tum14} with direct measurements. 

Then, the second key issue concerns D/H measurements. Today it rests on the measurements of \citet{Coo18}. The primitive abundance of deuterium can be determined from the observation of DI and HI lines from neutral clouds (Damped Lyman-$\alpha$ systems, DLAs) at high redshift, located on the line of sight to background quasars. While progress has been done to obtain precise measurements, these remain very scarce. Because of this, each measurement has therefore an important impact on the determination of the primitive D/H abundance (i.e. the mean value) and its accuracy must be tested intensively. Indeed, both values and associated uncertainties remain debated (e.g. the remeasurement of the deuterium abundance at $z=3.256$ by \citet{Riemer-Sorensen:2014aoa}).  More observations are crucially needed not only to decrease statistical errors but also have the potential to reveal subtle systematics. While several thousands DLAs have been detected thanks to large spectroscopic surveys (e.g. \citet{Noterdaeme:2012gi}), most of the background quasars are too faint for efficient selection of follow-up targets and precision measurements with current telescopes.  Notwithstanding, there is still some room to detect new bright quasars and hence potentially useful DLAs. For example, the QUBRICS bright Quasar survey has recently identified 55 new high redshift quasars ($z > 2.5$) \citep{Cal19,Bou20}. Alternatively, high-resolution optical spectrographs on the next generation of 30-m class telescopes will increase the number of accessible quasars and automatically the number of targets suitable for measuring D/H. For example, HIRES on the Extremely Large Telescope could increase the precision to 0.3\% with a five-fold increase in sample size, provided its wavelength coverage extends enough to the blue (\citet{Mai13} and Pasquier Noterdaeme, private communication). 

With the existing D/H data~\citep{Coo18}, the updated nuclear network and the slight shift of the baryonic density determined by Planck-2018, we witness a $\simeq 1.8\sigma$-tension on the baryonic density between BBN and CMB+BAO or equivalently between the D/H abundance prediction assuming (CMB+BAO)-baryonic density and its spectroscopic measurement. This is indeed a mild warning but it sheds some light on the sector of the big-bang theory, indicating that it should be watched carefully, both on the nuclear and astrophysical data sides.

It is worth mentioning that the Hubble constant tension has been interpreted as an early/late universe tension, while it shall maybe be seen as a thin/large beam tension~\citep{Fleury:2013uqa,Fleury:2018cro,Fleury:2018odh}. This new emerging tension, to be confirmed by more BBN and large scale estimations of the baryonic density, is a primordial/late time tension, so that the CMB would be tied between two lever arms at redshifts of order $z\sim1$ and $z\sim10^8$. If confirmed, the status of BBN, with the lithium-problem and a mildly-constraining helium-4, would have to be reconsidered. Note also that unlike the cosmological lithium problem, this deuterium tension can be mitigated easily by invoking a small contribution from most models developed to solve the lithium problem as they overproduce deuterium~\citep{AV12,Oli12,Coc:2014gia,Kus14,Coc15}. To finish, note
also that since BBN theory assumes a perfect FL geometry and since the spectroscopic data are located on our past light-cone at low redshift -- and thus well inside the CMB sky -- the Copernican principle could be at stake~\citep{Regis:2012iq,Dunsby:2010ts}. 

With this new data, cosmology shows once more that precision cosmology should come with a cosmology of correctness~\citep{Uzan:2016wji} and that the new tensions we witness are some precursory signs of a more realistic model or just a transient that would disappear with future data with better accuracy and better controlled systematics.

\section*{Note added} After this paper was submitted, two papers
addressing the same topic as our current work were posted \citep{Pis20,Yeh20}, confirming that differing conclusions can be traced to the data selection and analysis of \ddn\ and \ddp\ rates.

%%%%%%%%%%%%%%%%%%%%%%%%%%%%
\section*{Data availability}
%%%%%%%%%%%%%%%%%%%%%%%%%%%%

There are no new data associated with this article. The BBN code {\tt PRIMAT} is freely available at~\url{http://www2.iap.fr/users/pitrou/primat.htm}.

%%%%%%%%%%%%%%%%%%%%%%%%%%%%
\section*{Acknowledgements}
%%%%%%%%%%%%%%%%%%%%%%%%%%%%

We thank warmly Pasquier Noterdaeme for discussions on the detection of deuterium, Silvia Galli for her help with the baryon abundance posteriors from the Planck-2018 results, Ofelia Pisanti for discussions concerning the LUNA rates, Julien Froustey for careful reading of the draft, and Christian Iliadis for a long time collaboration on reaction rates evaluations.

%%%%%%%%%%%%%%%%
\section*{Erratum}
%%%%%%%%%%%%%%%%

We take the opportunity of the new release of the BBN code {\tt PRIMAT} to correct a few minor typos in \citet{Pitrou2018PhysRept}.
\begin{itemize}

\item  In Eq.~(53), there is an additional minus sign in front of $0.01452$.

\item In Eq.~(61), the function ${\cal S}$ must be multiplied by the constant factor $\bar s_\gamma$, both in the numerator and the denominator.

\item In Eq.~(105), the numerical value for $\lambda^{{\rm RC} 0}_0$ should be $1.75838$.
  
\item In Table 5, the first value in the RC+FM+WM+ID line (and which corresponds to $Y_{\rm P}$) should be 0.24710 instead of 0.24720.
  
\item In Table 7, 10$^{10}$ should be read instead of 10$^{5}$ for \sep\ multiplicative prefactor.

\item In Table 8, theoretical abundance values and errors should be as reported in Table \ref{TableAbundances}.

\item In Eq.~(B23), one must first read $g_\nu^{(2,0)}$ instead of $g_\nu$ in the first term, and in the fifth line $m_{\rm n}/m_{\rm p}$ must be replaced by $(m_{\rm n}/m_{\rm p})^{\pm 1}$.

\end{itemize}

These typos are corrected in the arXiv version of \citet{Pitrou2018PhysRept}.

%%%%%%%%%%%%bibliography

\bibliographystyle{mn2e}
%\interlinepenalty=10000
\bibliography{BiblioBBN}

\label{lastpage}
\end{document}